# Z-scan confocal method for indirect focusing location

P. Castro-Marín [1], G. Castro-Olvera [1], C. Ruíz [2], J. Garduño-Mejía [1], M. Rosete-Aguilar [1], and N. C. Bruce [1]

[1] *Centro de Ciencias Aplicadas y Desarrollo Tecnológico, Universidad Nacional Autónoma de México, Av. Universidad 3000, Coyoacán, Distrito Federal 04510, México.*
[2] *Instituto Universitario de Física Fundamental y Matematicas and Departamento de Didactica de la Matematica y de las ciencias experimentales, Universidad de Salamanca, Patio de las Escuelas s/n, Salamanca, Spain*

**Abstract** We present a new technique that we have defined as the z-scan confocal method to determine the absolute location and size of the focal spot in a tight focused ultrashort laser pulse. The method permits to accurately position a target in the focal spot with a fast response. The technique is designed to help to automatize the location of an overdense target in focus in a laser plasma experiment. The method allows for a fast localization of the focal position and the relative motion of the target with respect to it. As an example of the capacities we measure the defocusing of a fast rotating disc in several radius to reconstruct the motion of the disc at focus.

**I. INTRODUCTION**

The availability of ultrahigh high laser systems has opened new perspectives for applications. Most importantly, perhaps, are the applications of laser driven plasma accelerators which can produce high energy particles in very short distances and very short times. Ultrashort pulses exceeding intensities of $10^{18}$ W/cm$^2$ can create a plasma and accelerate electrons with electric gradients up to 3 or 4 order of magnitude larger than conventional accelerators. Interaction with underdense plasmas can produce electrons with high energies (>1 GeV) [1] which can in turn produce X rays and gamma beams with high brightness through different experimental schemes [2]. Solid targets are also used to produce proton and ion beams [3,4], x rays and gamma beams [5]. In most of the schemes, electrons in the overdense plasma can accelerate ions as they leave the target in the rear side to produce high energy broadband ions. All these sources have remarkable properties and many proof of principles have been developed to demonstrate the physics of these new sources. In order to produce a bigger impact in applications, these sources need to demonstrate larger average powers which means to increase the repetition rate as which they are produced.

This scaling to high repetition rates in the case of solid targets need the development of complex targets technologies and solutions to produce these plasma sources with good stability and repeatability. Also in the case of solid targets, tight focusing is needed and that increase the complexity of precise alignment since the associated Rayleigh length is very small and deviations from the focal plane reduce the intensity significantly. The alignment is even more complicated as the repetition rate increase because the target needs to be renovated and the alignment needs to be measured and corrected. Therefore, accurate, simple and fast techniques to locate the focal point are needed to solve this problem.

Several techniques are available for this purpose. Recently the CHIP technique [6] has been developed which is able to perform micron alignment in nanostructured targets [7]. Other techniques use backlight illumination [8] or three colors interferometry [9]. These techniques are good in some circumstances but can only position the target but not characterize the laser focal spot. This might be important in situations where coupling between spatial and temporal parts of the pulse are important.

In this paper we present a new technique called the z-scan confocal method for indirect focusing location to characterize the focal relative position respect a solid target, at the confocal plane with micron accuracy. The method allow a quantitative measure of the defocusing, which can be later corrected. Due the technique use a fast detector, the method is very well suited for high frequency targets.

1. **The z-scan confocal method for indirect focusing location and general considerations.**

The concept of the z-scan confocal method for indirect focusing location is an extension to a method previously introduced: Pinhole Masked Linear z-scan (PML z-scan) [10]. The main idea is that the amount of focused light passing through a narrow aperture, is maximized at the focal plane and this location can be measured indirectly, at the confocal plane, with a parallel second detection stage. Once the focal plane is located, we can place a solid target, instead of the detection setup, and use the second stage, at the confocal plane, to scan again and positioning the target at the focal plane or map the defocusing of a rotating target as we demonstrate in the applications section.

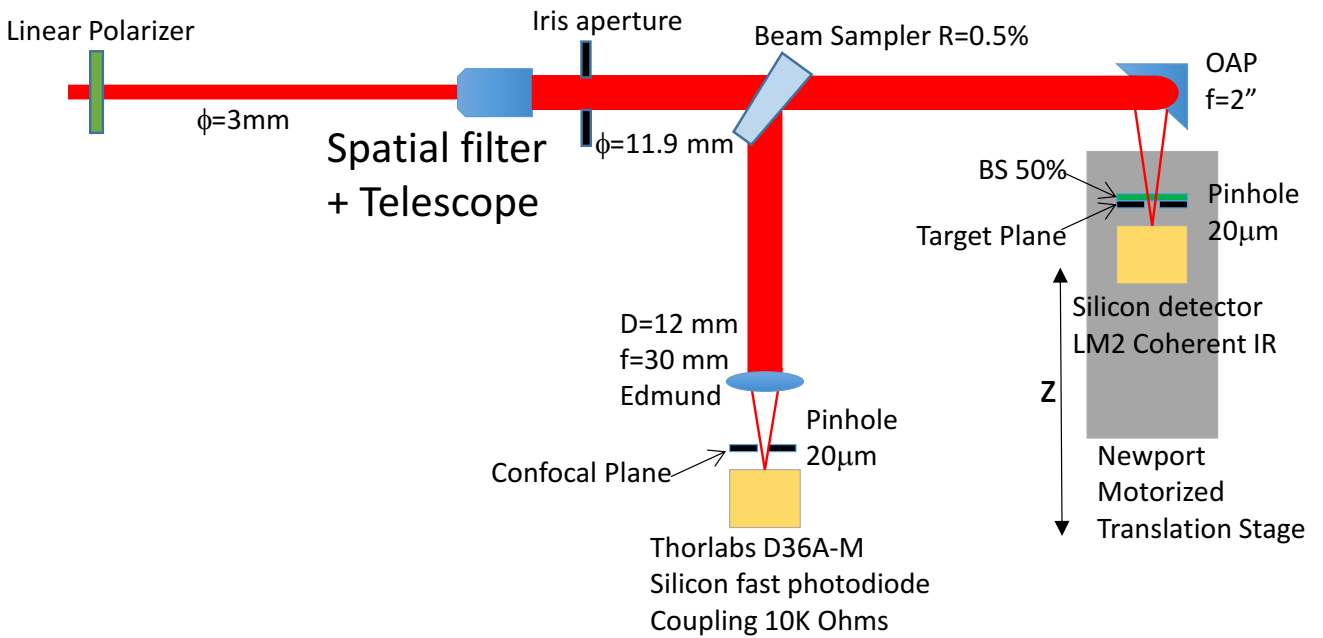

Figure 1. The z-scan confocal method for indirect focusing location

The laser test system consist on a home-made femtosecond Ti:Sapphire oscillator which delivers near bandwidth limited pulses (BL) of 58 fs (FWHM) with central wavelength @ 808 nm, 30 nm broadband, with 1.1 W average power, 10 nJ per pulse @ 112 Mhz, repetition rate. The laser beam is transmitted through a linear polarizer followed of a telescope and a spatial filter, to obtain a collimated beam with 11.9 mm diameter (figure 1). After the telescope, the beam is transmitted through a 0.5 % reflectance beam sampler (BSF10-B1, Thorlabs). The main part of the beam is focused with an 90º Off Axis Parabolic silver coated mirror (OAP) of 2 inches focal length (MPD129-P01, Thorlabs). In the experimental setup we have defined two focal positions, at the target and at the confocal plane. As a first step, we have verified the simultaneous detection of the focal planes locations at each side.

At the target position, a thin 1 mm partially Au coated (R=50%) objective slide (BS) is attached on a 20 um pinhole (P20S, Thorlabs). Right after the pinhole a silicon optical sensor (Coherent, LM2-VIS), measures the transmission through the pinhole. The BS, the pinhole and the optical sensor are mounted on a motorized translation stage (Newport M-UTM50CC1DD) with resolution up to 1 um is used to scan along the z direction. The scan produces a signal of z vs. amplitude which maximum defines the location of the focal plane. A shift in the position of the maximum is expected due the thickness of the BS. The width of the signal is related to the Rayleigh range length of the OAP (target position). To prevent saturation, in the silicon photodiode, LM2-VIS, we have used a linear polarizer as a filter to control the laser input power.

At the confocal section, the second detection stage, the partially reflected beam from the BS surface is collimated back to the beam sampler and deviated at 90 degrees towards a lens with focal length of 30 mm (Achromatic doublet for NIR model NT45-794, Edmund). Focused light pass through a 20 um Pinhole, attached in front of a high-speed silicon detector (14 ns rise time, Thorlabs DET36A-M) coupled to a 10K Ohms resistance, with the objective to increase the sensitivity. If the target is out of focus then, the reflected back beam will not be any longer collimated, and then the amplitude of the reflected light into the fast photodiode, will be reduced and the opposite. The location of the maximum of the signal, with respect to z, measured indirectly at the confocal stage, will define the focal plane position of the target. Actual experimental setup is presented in figure 2.

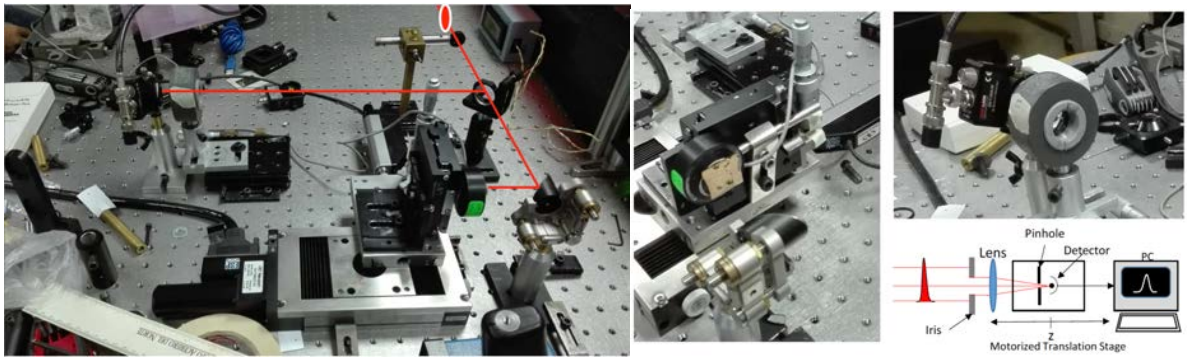

Figure 2. Experimental setup pictures

Once both signals, at the reflecting surface in the target plane and confocal position are acquired simultaneously, a solid target is placed at the focal plane in the detection stage. In our case we have used a hard drive disc as a target. The procedure is described in the following section.

**B. Results**

In order to validate and demonstrate the confocal z-scan method for indirect focal position location technique we measure the focalization of a short laser pulse in a fast rotating disc from a computer hard drive. In Figure 3, we present the result from the simultaneous detection at both conjugate and target positions, using 20 um pinholes in each detection stage.

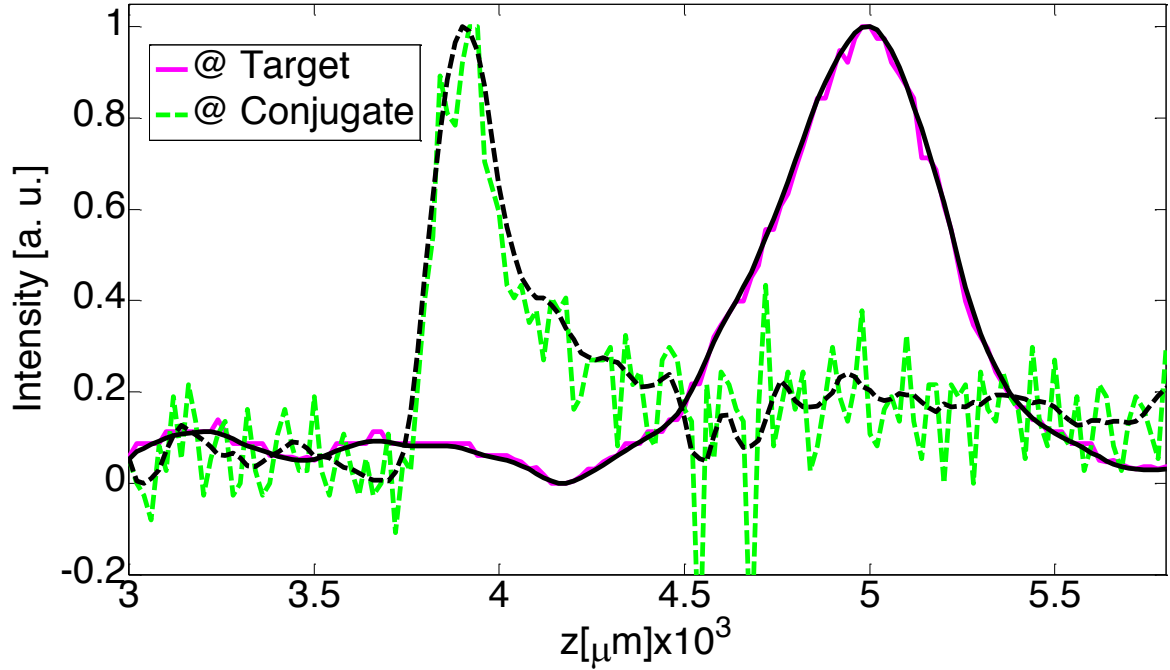

Figure 3. Indirect focal position location at target position with the confocal z-scan method. Simultaneous signals z vs. amplitude are detected at the target (solid lines) and the conjugated position (dashed lines). Black lines corresponds to the signals after a filtering process.

The relative separation between the two signals are attributed to the thickness of the partially reflective Au coated microscope slide (BS), located at the target position.

In both cases, a smoothing process was performed to filter out part of the noise. By adjusting the filtered signals with a Gaussian fit, at the target and confocal position we have obtained the FWHM at both locations, these are 574 um and 207 um respectively. For the last one, an exclusion rule was applied in the nonsymmetrical part of the trace. The relative separation between the two maximums is about 1100 nm, which is consistent with the BS thickness (microscope objective slide). The actual measurement of the power, at the peak of the signal, corresponds to about 1 mW at the target position and 100 nW at the conjugate detection stage, which represents about $10^{-4}$ ratio demonstrating the sensitivity of the technique. The actual defocus position, at the maximum intensity peak value, can be located very accurately with an error of a couple of microns.

It can be seen from Figure 3 that the target scan is symmetric whereas the conjugate position scan is not. The target position scan is symmetric because this is simply a map of the through focus intensity distribution which, close to focus, is symmetric [11]. However, the scan in the conjugate plane is not symmetric. The intensity on the detector depends on the object position for the conjugate imaging lens, which, with the lens focal length, determines the position of the image, and the conjugate image position relative to the detector position determines the signal. Now, the object position for the conjugate imaging lens is the position of the image of the light reflected by the microscope slide and focused by the OAP mirror. But this image position depends, through the Gauss equation on the position of the object, or focussed point of light after the first pass of the OAP mirror. Thus if the target position is before focus, the object point is at a smaller distance than the focal length of the mirror, and the image point is virtual, whereas if the light is focussed before the target, the object distance is larger than the focal distance, and the image is real. This difference in image position gives the asymmetry in the trace (conjugate).

**Experimental results of hard disc drive (HDD) used as a target**

After the validation of the technique, the detection stage at the target position was replaced with a Hard Disc Drive (HDD), which has been used as a rotating solid target. The rotation was controlled externally using a microprocessor with amplification stages, generating accurate control but limited to low frequencies. The alignment of the HDD and its location at the focal plane was achieved by measuring at the confocal plane. After this, the focal tracking system was tested by measuring in three different points at the HDD surface (see figure 4).

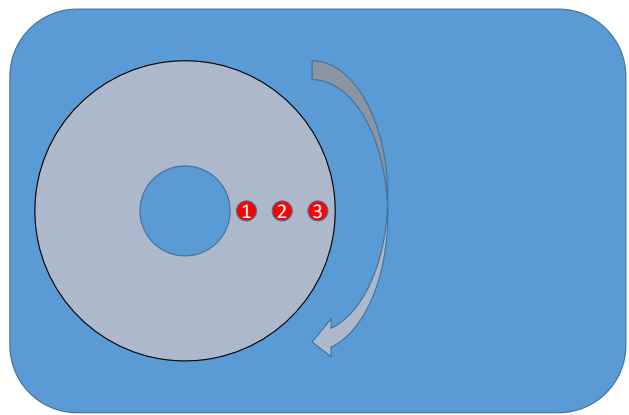

Figure 4. HDD schematics

The following results demonstrate the detection of the precession movement of the HDD near the center, middle position and near the edge. Detection was performed at the confocal position, and the signal is analyzed in both time and the frequency domain. Actual experimental results are presented for the three different locations at HDD surface (target): figure 5 (near the center position, red mark 1), figure 6 (near the middle position, red mark 2) and figure 7 (near the edge position, red mark 3).

In the figures, the dots corresponds to actual experimental measurements, (Signal) and the line correspond to a signal processing and it reconstruction (Analysis).

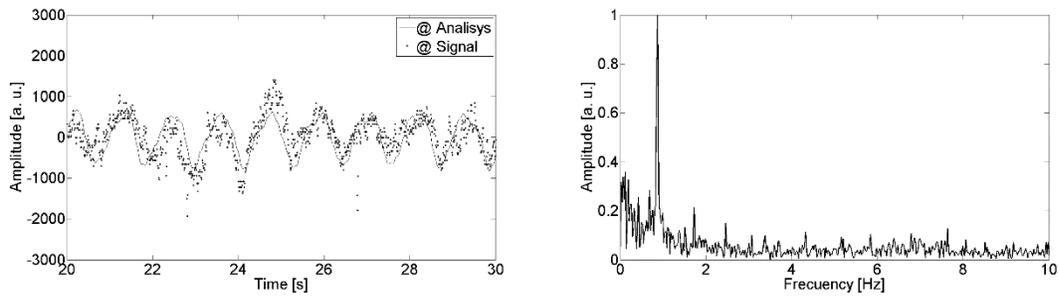

Figure 5. Experimental signal measured near the center position on the HDD.

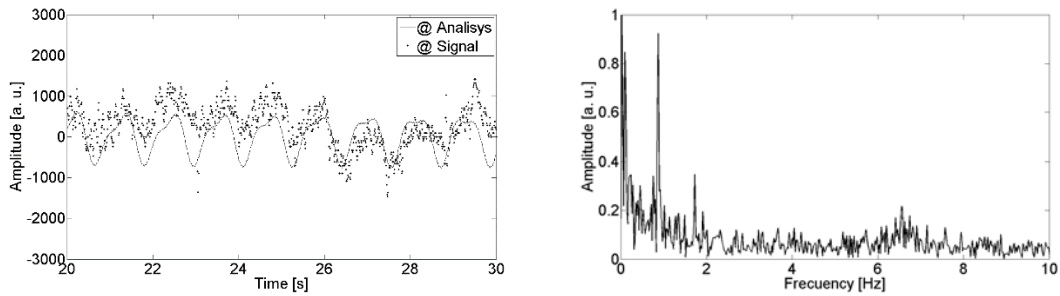

Figure 6. Experimental signal measured near the middle position on the HDD

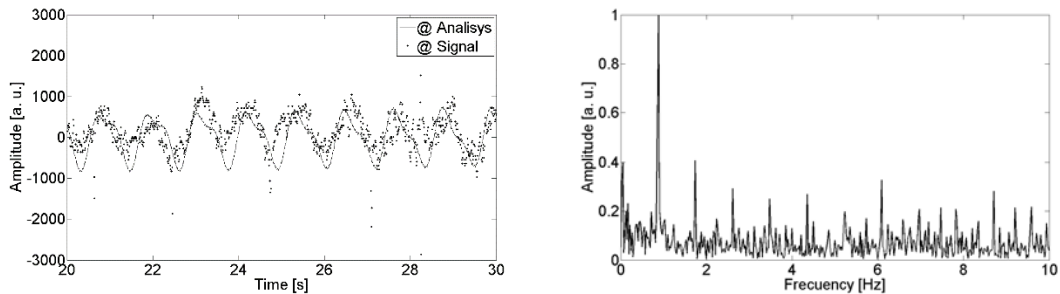

Figure 7. Experimental signal measured near the edge position on the HDD

From these results we can measure an actual fundamental angular frequency of 0.87 Hz on the three cases. In the frequency domain is clear that the movement presents harmonic components. Such of harmonic frequencies can be associated with small vibrations, which becomes more evident on the regions further away from HDD rotation support mechanism.

The procedure was repeated at 1.05 Hz to demonstrate the accuracy in the detection (see figure 8).

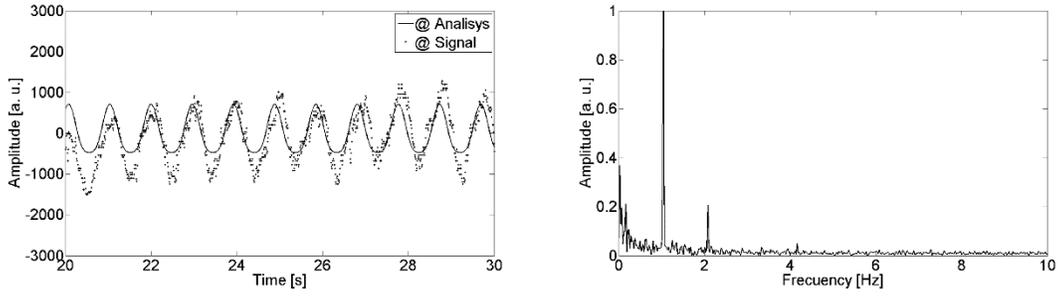

Figure 8. Experimental signal measured near the edge position on the HDD at 1.05 Hz angular frequency

For the signal processing, a noise filtering and signal reconstruction with a Taylor Series method has been performed. On this process, the DC component in the frequency domain is removed by subtracting the offset in the time domain. After this the fundamental and the harmonic frequencies are located ($f_i$) with its corresponding amplitudes ($a_i$), figure 9, and signal in the time domain, S(t), is retrieved.

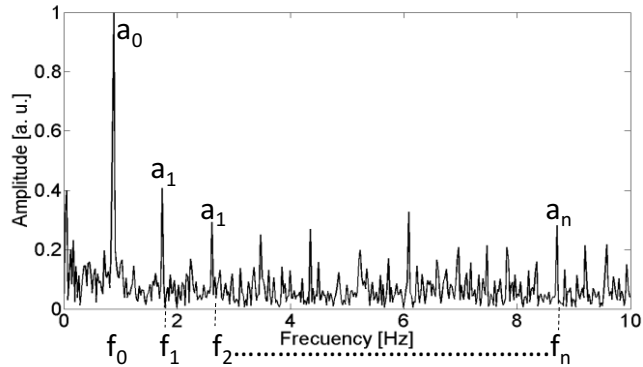

Figure 9. Frequency location and its amplitudes

For signal reconstruction, a Taylor Series is applied according the following equation 1:

$$S(t) = \sum_{n=0}^{N} a_n \cos(2\pi f_n t + \phi) \qquad [1]$$

In the next figures we present the signal reconstruction in the time domain: figure 10 (near the center position), figure 11 (near the middle position) and figure 12 (near the edge position).

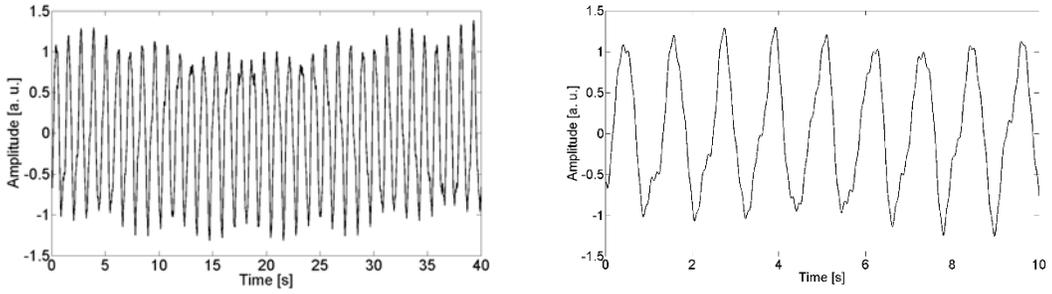

Figure 10. Signal reconstruction near the center position (full span and zoom scale)

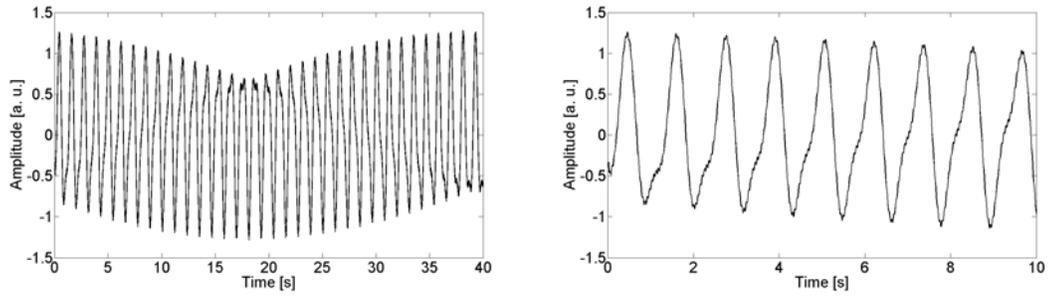

Figure 11. Signal reconstruction near the middle position (full span and zoom scale)

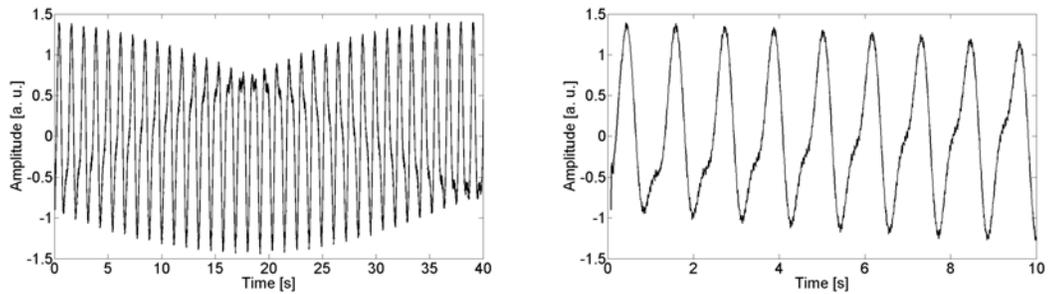

Figure 12. Signal reconstruction near the edge position (full span and zoom scale)

From these figures, we can observe the harmonic frequencies effect on the actual signal in the time domain. It demonstrate a modulation in the signal, which becomes more evident at the edge of the surface of the target. From these results we can observe that the modulation strength also depends on the distance respect the center rotation of the HDD. The periodic modulation is demonstrated in the zoomed figures.

**Conclusion**

In this work we present a novel technique that we have defined as the z-scan confocal method for indirect focusing location to characterize the focal relative position respect a solid target, measured indirectly at the confocal plane. The technique demonstrates high sensitivity of about tens of nW, in terms of required optical power at the detection stage, and spatial resolution in the order of couple of microns to locate the focal plane. Sensitivity of the technique allows detecting periodical focal shifts and vibrations of a rotating solid target, as a function of the distance, respect the rotation center. Due the technique use a fast detector, the method is very well suited for the focal tracking on high frequency rotating solid targets.

Funding

Dirección General de Asuntos del Personal Académico, Universidad Nacional Autónoma de México (DGAPA, UNAM) (PAPIIT-IG100615). Camilo Ruiz also thanks MINECO SPAIN project FIS2016-75652-P

Acknowledgments

Pablo Castro-Marín and Gustavo Castro-Olvera acknowledge a grant to Consejo Nacional de Ciencia y Tecnología, CONACyT, México and to the Programa de Maestría y Doctorado en Ingeniería, UNAM.

References

[1] E. Esarey, C. B. Schroeder, and W. P. Leemans, Physics of laser-driven plasma-based electron accelerators, Rev. Mod. Phys. 81, 1229 (2009)

[2] S. Corde, K. Ta Phuoc, G. Lambert, R. Fitour, V. Malka, A. Rousse, A. Beck, and E. Lefebvre, "Femtosecond x rays from laser-plasma accelerators", Rev. Mod. Phys. 85, 1 (2013)

[3] A. Macchi, M. Borghesi, M. Passoni, "Ion acceleration by superintense laser-plasma interaction", Rev. Mod. Phys **85**, 751-793 (2013)

[4] H. Daido, M. Nishiuchi, A. S. Pirozhkov, "Review of laser-driven ion sources and their applications", Reports on Progress in Physics, Volume 75, Number 5 (2012).

[5] C. Courtois, R. Edwards, A. Compant La Fontaine, C. Aedy, M. Barbotin, S. Bazzoli, L. Biddle, D. Brebion, J. L. Bourgade,1 D. Drew, M. Fox, M. Gardner, J. Gazave, J. M. Lagrange, O. Landoas, L. Le Dain, E. Lefebvre, D. Mastrosimone, N. Pichoff, G. Pien, M. Ramsay, A. Simons, N. Sircombe, C. Stoeckl, and K. Thorp, "High-resolution multi-MeV x-ray radiography using relativistic laser-solid interaction", Physics of Plasmas **18**, 023101 (2011)

[6] C. Willis, P. L. Poole, K. U. Akli, D. W. Schumacher, and R. R. Freeman "A confocal microscope position sensor for micron-scale target alignment in ultra-intense laser-matter experiments." *Review of Scientific Instruments* 86.5 (2015): 053303.

[7] M. Blanco Fraga1, M. Flores Arias, C. Ruiz Mendez and M. Vranic, "Table-top laser-based proton acceleration in nanostructured targets", 2017 *New J. Phys.* **19** 033004

[8] D.C. Carroll and P. McKenna, S. Kar and M. Borghesi  P. Foster, D. Symes, R. Pattathil and D. Neely,  "An imaging system for accurate target positioning for fast focusing geometries", CLF Annual Report, 2011-2012

[9] N. Booth, O. Ettlinger, D. Neely, R. Pattathil, A. Sellers and D. Symes, Sub-micron accuracy target alignment Central Laser Facility, STFC Rutherford Appleton Laboratory, Harwell Oxford, Chilton, Didcot, Oxon, OX11 0QX, UK in the workshop  Targ1: Targetry for Laser-driven Proton (Ion) Accelerator Sources

[10] P. Castro-Marín, J. Garduño-Mejía, M.Rosete-Aguilar, N. C. Bruce, D.T. Reid, C. Farrell, G. E. Sandoval-Romero, "Aberration analysis based on pinhole-z-scan method near the focal point of refractive systems", Optical Modeling and Performance Predictions VIII, Proc. of SPIE Vol. 9953, 99530Q-1, 2016,  doi: 10.1117/12.2238239

[11] M. Born and E. Wolf, "Principles of Optics", Cambridge University Press, 1999